\documentstyle[12pt,aaspp,tighten]{article}

\def\cJ{{\cal J}}
\def\kms{{\rm\,km\,s^{-1}}}

\def\kpc{{\rm\,kpc}}

\def\sgn{\mathop{\rm sgn}\nolimits}

\begin{document}

\title{Kinematic Signature of a Rotating Bar Near a Resonance}
\author{Martin D. Weinberg}
\affil{Department of Physics and Astronomy\\University of
Massachusetts, Amherst, MA \ 01003}

\begin{abstract}
There have been several recent suggestions that the Milky Way has
rotating bar-like features based on HI and star count data.  In this
paper, I show that such features cause distinctive stellar kinematic
signatures near OLR and ILR.  The effects of these resonances may be
observable far from the peak density of the pattern and relatively
nearby the solar position.  The details of the kinematic signatures
depend on the evolutionary history of the `bar' and therefore velocity
data, both systemic and velocity dispersion, may be used to probe the
evolutionary history as well as the present state of the Galaxy.

Kinematic models for a variety of simple scenarios are presented.
Models with evolving pattern speeds show significantly stronger
dispersion signatures than those with static pattern speeds,
suggesting that useful observational constraints are possible.  The
models are applied to the proposed rotating spheroid and bar models;
we find: 1) none of these models chosen to represent the proposed
large-scale rotating spheroid are consistent with the stellar
kinematics; and 2) a Galactic bar with semimajor axis of $3\kpc$ will
cause a large increase in velocity dispersion in the vicinity of OLR
($\sim5\kpc$) with little change in the net radial motion and such a
signature is suggested by K-giant velocity data.  Potential future
observations and analyses are discussed.

\end{abstract}

\keywords{galaxy: kinematics, galaxy: structure,  stellar dynamics}

\section{Introduction}

Recently, several groups have suggested that the Milky Way may have
one or more non-axisymmetric structures, such as stellar bars and
tri\-axial spher\-oids (e.g. \cite{bli91}, hereafter BS;
\cite{wei92}).\footnote{Throughout, I will call the general class of
bisymmetric rotating disturbances `bars'.} The presence of a bar was
inferred from an analysis of IRAS source counts, while gas kinematics
were used to deduce the existence and structure of a triaxial spheroid.
In either case, the relationship between the non-axisymmetric density
structure and the associated kinematic effects are well defined by
theory.  Briefly, a structure with a rotating pattern (as has been
suggested both for the BS spheroid and for the bar), in a disk system
with nearly circular orbits will generate three strong resonances: the
inner Lindblad resonance (ILR), the outer Lindblad resonance (OLR),
and a resonance at the location of corotation (CR).  Orbits between
ILR and CR will be elongated in a direction parallel to the bar while
orbits outside OLR will be elongated perpendicular to the bar.

However, the impact of a bar on stellar orbits within a disk is not
fully described by standard epicyclic theory particularly near
resonances, where nonlinear effects may cause discontinuous changes in
orbit morphology and trap apoapses into a narrow range of position
angles.  Furthermore, the standard theory assumes the existence of a
static bar, whereas the kinematic consequences of a bar are
inextricably linked to its past history.  This paper shows that
resonances between disk and bar potentials produce strong kinematic
signatures.  The simplest case is that where the bar grows
adiabatically.  The case of a bar with an evolving pattern speed is
also considered.  The results show that the integrated effect of
bar-disk resonances on stellar kinematics allow the qualitative
features of the evolution to be diagnosed.

The general approach is described in \S2.  The response near resonance
is approximated by a non-linear Hamiltonian model which is exact to
first order in epicyclic amplitude.  The qualitative behavior of
orbits near resonance is explored.  Some readers may wish to skim \S2
and proceed to \S3, where the models are applied to an ensemble of
orbits, and expressions for line of sight velocity and velocity
dispersions are derived.  In \S4, these results are applied to the
problems of a rotating spheroid and Galactic bar.  Existing kinematic
data are inconsistent with any of the rotating spheroid models
developed here, but are in remarkable agreement with a bar ending at 3
kpc.  Several suggestions for additional observations and analysis are
offered in \S4.3 and the results are summarized in \S5.

\section{Theory}

If the force due to a rotating bar is a small perturbation in the
region of interest, e.g., less than roughly 5\% of the axisymmetric
force, then the dominant response will be near resonances.  Standard
epicyclic theory describes the motion near one of these locations as a
coupled driven simple harmonic oscillator (SHO), whose amplitude
diverges near resonance.  In reality, however, non-linear terms detune
the oscillator so that it is no longer a SHO.  This section seeks a
more appropriate model for the motion near resonance.  Such a model
must take into account that the effect of the resonance depends on the
time history of the non-axisymmetric potential.  As the amplitude and
frequency of a rotating bar changes, its resonances sweep through
phase space.  As orbits are captured into libration and/or pass
through resonance, they may significantly change in epicyclic
amplitude and guiding center trajectory.  For these reasons, the
models presented below explicitly treat the bar potential in its
time-dependent form.

\subsection{Dynamical model}

Let us begin with a flat axisymmetric galactic disk whose dynamics are
then specified by the Hamiltonian function $H_o({\bf I})$ where $\bf
I$ are the actions.  The action-angle variables are the natural choice
to describe regular periodic motion (e.g. Goldstein 1950, Chap. 9).
One action may be identified with the angular momentum and the other
with the momentum of the radial motion.  Expanding the bar
perturbation in polar harmonics we may write
\begin{equation} \label{hamilt0}
H = H_o + \sum_{m=-\infty}^{\infty}U_m(r)\exp\left[im\left(\phi -
\int\Omega_b\,dt\right)\right].
\end{equation}
The quantity $\Omega_b$ is the pattern speed of the bar which may
explicitly depend on time.
Since the unperturbed Hamiltonian is cyclic in the conjugate angles,
it proves convenient to expand the terms of the sum in equation
(\theequation) as a Fourier series in the unperturbed actions $\bf I$
and angles $\bf w$.  In addition, since a bar-like perturbation is
likely to be dominated by the quadrupole, let us restrict the sum in
equation (\theequation) to a single term ($m=2$); other multipole
terms may be treated similarly.  We may write (c.f.
\cite{lyn72,tre84})
\begin{eqnarray} \label{actionexp}
U_m(r)\exp\left[im\left(\phi-\int\Omega_b\,dt\right)\right] &=& \nonumber \\
	&&\hskip-50pt \sum_{l_1=-\infty}^\infty \sum_{l_2=-\infty}^\infty
	W^m_{l_1\,l_2}(I_1, I_2) \exp\left[i\left(l_1w_1 + l_2w_2 -
	m\int dt\,\Omega_b\right)\right],\\
\noalign{\hbox to\hsize{where\hfill}} \label{expcoef}
	W^m_{l_1\,l_2}(I_1, I_2) &=& {1\over\pi}\int^\pi_0 dw_1\,
	\cos[l_1w_1 - l_2(\phi-w_2)]U_m(r).
\end{eqnarray}
The right-hand-side of equation (\ref{actionexp}) is a Fourier series
in the angle variables $w_j$ whose Fourier coefficients
$W^m_{l_1\,l_2}(I_1, I_2)$ are given by equation (\ref{expcoef}).
Furthermore, since the Galactic disk is rotationally supported with
relatively small radial excursion, we may represent the unperturbed
orbits in the epicyclic limit.  For definiteness, we take the rotation
curve to be flat with rotation velocity $V_{rot}$.  With these limits,
the orbital quantities and the action-angle variables become:
\begin{eqnarray}
	r &=& R + a\sin w_1,				\label{actone-1} \\
	I_1 &=& {V_{rot}\over\sqrt{2}}{a^2\over R},	\label{actone} \\
	I_2 &=& V_{rot} R,				\label{acttwo} \\
	\Omega_1 &=& {\sqrt{2} V_{rot}\over R},	\\
	\Omega_2 &=& {V_{rot}\over R},		\\
	w_1 &=& \Omega_1 t + w_{10},			\label{angone} \\
	w_2 &=& \phi - {\Omega_2\over\Omega_1}{2a\over R}\cos w_1,
							\label{angtwo}
\end{eqnarray}
where quantities $I_1$ and $I_2$ are the radial and azimuthal actions
with corresponding angles $w_j$ and frequencies $\Omega_j$, $R$ is the
guiding center of the trajectory and $w_{10}$ describes the radial
phase at $t=0$.  Using the above relations, we may now explicitly
evaluate equation (\ref{expcoef}):
\begin{eqnarray} \label{epiexp}
	W^2_{l_1\,l_2} &=& \delta_{l_2\,2}\exp[-im\Omega_b t]\exp[-i\pi/2]
		{a\over R} \times \nonumber\\
		&&\quad\left\{
		\delta_{l_1\,0} {2\Omega_2\over\Omega_1} U_2(R) +
		\delta_{|l_1|\,1} \sgn(l_1)
		{R\over2}\left.{dU_2\over dr}\right|_{R}\right\} \nonumber\\
		&&\quad + c.c. + {\cal O}(a^2),
\end{eqnarray}
where $\delta_{ij}$ is the Kronecker delta.
The epicyclic approximation is consistent as long as the epicyclic
amplitude $a$ remains significantly smaller than $R$ throughout the
evolution and this condition must be checked explicitly.  If desired,
equation (\theequation) may be explicitly rewritten in terms of the
actions using equations (\ref{actone-1})--(\ref{angtwo}).
Substituting into equation (\ref{hamilt0}) and (\ref{actionexp})
yields the governing Hamiltonian in terms of the action-angle
variables.

We are interested in the effects near a particular resonance defined by
a commensurability between frequencies:
\begin{equation} \label{rescon}
l_1\Omega_1 + l_2\Omega_2 - m\Omega_b=0.
\end{equation}
It follows that the angle $w_s\equiv l_1w_1 + l_2w_2 - m\int
dt\,\Omega_b$ is very slowly varying near resonance.  To make the
evolution near resonance explicit, we can rewrite the Hamiltonian
(eq. \ref{hamilt0}) with $w_s$ as one angle.  To do this, we may
define the generating function of the canonical transformation (2nd
kind, \cite{gol50}) as follows:
\begin{equation} \label{genS}
	S = \left(l_1w_1 + l_2w_2 - m\int dt\,\Omega_b\right)J_s + w_1
	J_f,
\end{equation}
which gives the following new set of variables:
\begin{eqnarray}
	J_f &=& I_1 - {l_1\over l_2}I_2,	\label{slowfast} \\
	J_s &=& {1\over l_2} I_2,		\label{slowfast+1} \\
	w_f &=& w_1,				\label{slowfast+2} \\
	w_s &=& l_1w_1 + l_2w_2 - \int dt\,\Omega_b,
						\label{slowfast+3}
\end{eqnarray}
Since $w_s$ is slowly changing relative to the second ``fast'' angle
near resonance, relative to the timescale of the slow variable, the
motion in the fast variable is adiabatic.  Physically, the slow
variable correspond to the precession of the orbit in the frame of
the rotating disturbance and the fast variable correspond to the
motion of the particle around its orbit (e.g.  \cite{lyn72}).  We
exploit the disparity in slow and fast frequency to average over the
motion in $w_f$ and rewrite the Hamiltonian in terms of the slow
variables alone.  This is the so-called ``averaging principle'' (e.g.
\cite{Arnl:80,lic83}).

We may derive a simplified model describing near-resonant behavior by
expanding the averaged Hamiltonian, ${\bar H}$, about the value of the
slow action at resonance:
\begin{eqnarray} \label{hamexp}
	{\bar H} &=& \left[H_o(J_{s,res}) -
m\Omega_b(t)J_{s,res}\right]  +
\left.\left[l_1\Omega_1 + l_2\Omega_2 -
m\Omega_b(t)\right]\right|_{res}\left(J_s -
J_{s,res}\right) + \nonumber\\
&&\quad {1\over2}\left.{\partial^2H_o\over\partial J^2_s}\right|_{res}
\left(J_s - J_{s,res}\right)^2 +
\left.W^2_{l_1\,l_2}\right|_{res}\cos(w_s+w_r),
\end{eqnarray}
where the subscripts ``res'' denote values at resonance.  In equation
(\theequation), the first term on the RHS is independent of $J_s$ and
$w_s$ and the second term is zero at the resonance.  In general
$\Omega_b$ is a function of time, as indicated.  Therefore, if we
define $t_{res}$ to be the time at which the resonance occurs,
$l_1\Omega_1 + l_2\Omega_2 - m\Omega_b(t_{res})=0$, then equation
(\theequation) becomes
\begin{eqnarray} \label{hamres}
	{\bar H} &=& \left[H_o(J_{s,res}) - m\Omega_b(t)J_{s,res}\right]
	- m\left[\Omega_b(t)-\Omega_b(t_{res})\right]\left(J_s - J_{s,res}\right)
+\nonumber\\
&&\quad {1\over2}\left.{\partial^2H_o\over\partial J^2_s}\right|_{res}
\left(J_s - J_{s,res}\right)^2 + W^2_{l_1\,l_2}({\bf I})\cos(w_s+w_r).
\end{eqnarray}
We will further simplify the 2nd term by assuming that any time
dependence in $\Omega_b$ is slow compared to the evolution of a
particular orbit and write
$\Omega_b(t)-\Omega_b(t_{res})={\dot{\Omega}}_b(t-t_{res})$.

The character of the solutions to the Hamiltonian given by equation
(\theequation) depends on the functional behavior of $W^2_{l_1\,l_2}$
on $J_s$ (cf. eq. [\ref{expcoef}]).
If $l_1=0$, which gives the corotation resonance, then
$W^2_{l_1\,l_2}$ be will a constant to lowest order and equation
(\ref{hamexp}) then corresponds to the Hamiltonian of a simple
pendulum.  If $l_1=\pm1$, then $W^2_{l_1\,l_2}\propto
a\propto\sqrt{I_1}$.  This is not a pendulum but the qualitative
properties of the solution are similar.  Let us consider the
corotation case explicitly.  If the bar-like perturbation grows slowly
over time, then $W^2_{0\,2}$ slowly increases from zero.  Physically,
this corresponds to a pendulum with an increasing gravitational
constant (e.g. \cite{yod79}).  Initially, the pendulum will be
uniformly rotating around it pivot like a propeller.  The action, the
integral of the angular momentum around the trajectory, is an
adiabatic invariant and is conserved as the gravitational force grows.
However, if the gravity becomes strong enough the bob will not be able
to make it over the pivot.  At this transition point, the trajectory
has infinite period (the pendulum bob can just make it to the unstable
equilibrium in an infinite time) which breaks the adiabatic invariant.
After the transition, the motion is no longer rotation but swinging or
{\it librating} and the action is again conserved but with a different
value.  The new value of the action depends on the rate of growth at
the transition.  This transition point is often referred to as the
resonance, but the motion does not correspond to unbounded amplitude
in the sense of a forced SHO but zero frequency in the sense of
equation (\ref{rescon}).  The transitional or {\it critical}
trajectory divides the motion into regions of rotation and libration.

If $l_1=\pm1$, the qualitative properties of the solution are similar:
there can be both rotational and librational regions of phase-space
along with trajectories of infinite period giving rise to jumps in
action.

\subsection{Evolution Through Resonance: Example}

Although the dynamical model has been reduced to one degree of
freedom, and for example the motion may vanish in that dimension, the
orbit itself remains similar to a circular orbit.  The slow angle
describes the position of the apocenter relative to the bar and is
correlated with the variation of slow action in the presence of the
bar.  A change in slow action changes both the epicyclic amplitude and
guiding center radius.  As an example, Figure \ref{fig:trap} shows the
radius and angle of apocenter relative to the bar position angle,
$\phi_{apo}$, for an orbit trapped into libration at ILR.  The orbit
is initially a simple rosette with an r.m.s. radial velocity of 1/10
its tangential velocity.  As the bar strength grows, the orbit passes
through the critical trajectory which causes a large jump in the
epicyclic radius.  Notice that $\phi_{apo}$ is restricted to a narrow
range of position angles relative to the bar as expected for
libration.  Also, the angle of the apocenter is correlated with the
epicyclic amplitude and ``lingers'' near $\pm\pi/2$; the orbit is an
oval oriented orthogonal to the bar.  As $\phi_{apo}$ swings quickly
through zero, the orbit is either nearly circular or very elongated.
The plot of this orbit in the bar frame is shown in Figure
\ref{fig:traporb} (for simplicity only $1/4$ of the orbit shown in
Figure \ref{fig:trap} is displayed).  At $t\approx0$,
$\phi_{apo}\approx0$ and the orbit is rather eccentric.  The apocenter
angle quickly swings up to and lingers near $\pi/2$.  If the plot
continued, the orbit would appear more and more circular and the
apocenter would then rapidly swing to $\approx-\pi/2$ and become more
elongated, and then the cycle would repeat.  This orbit responds
strongly to the bar perturbation because it is very slowly precessing
in the bar frame.  The same orbit viewed from the inertial frame is
shown in Figure \ref{fig:intlorb}; it appears to be an unremarkable
rosette orbit.

Figures \ref{fig:untrap} and \ref{fig:untrpf} are similar to Figure
\ref{fig:trap} but for rotating orbits with initial guiding centers
at $R=1.0$ and 2.5 respectively.  The orbit at $R=1.0$ passed through
the critical trajectory and the one with $R=1.7$ did not; note the
large difference in the amplitude of radial oscillation.  As $R$
decreases, the orbit does cross the critical trajectory (at
$R\approx1.6$) and the amplitude of radial oscillation doubles.  The
existence of critical behavior causes a radially well-demarcated band
of resonance-induced kinematic changes in the stellar disk which may
be observed.  The graphs of these orbits in rotation are similar to a
rosette even in the bar's rotating frame.  However, the position angle
remains correlated with epicyclic amplitude even though $\phi_{apo}$ is
takes on all angles roughly evenly.  The amplitude is strongest near
$\phi_{apo}=0,\pi$ and causes the observed kinematics for orbits near
resonance to vary with respect to the bar position angle.

The location and size of the band where resonant effects are
significant depends on the history of the bar perturbation.  For
example, if the pattern speed changes as the bar evolves, the band may
be wider with a different fraction of librating to rotating orbits.
In fact, the relative fraction of librating orbits is very different
depending on the sign of ${\dot\Omega}_b$.  The mathematical details
of the model Hamiltonian (eqs. \ref{hamres}, \ref{hamlr}) and its
solution are discussed in Appendices A and B. Quantitative
observational predictions for a variety of Galactic scenarios are
discussed in \S3.

\section{Application to observed velocities}

In this section, we will describe the general features of the
kinematic signatures near resonance.  We begin by determining the
model from \S2 (and Appendix A) for two specific scenarios: the
response of the Galactic disk to a rotating Galactic bar near OLR and
to a rotating triaxial spheroid near ILR.  In both cases, the models
either have constant or decreasing pattern speeds and are chosen to
illustrate the sensitivity of the kinematic signatures to the history
of the evolution.  We will compare these models with Galactic
observations in \S\ref{sec:galaxy}

The Galactic rotation curve is assumed to be everywhere flat with
value $V_{rot}$. This is a fair approximation for radii of interest:
$R\gtrsim4\kpc$ (e.g. \cite{ScTe:83,FiBS:89}).  We take the potential of
the triaxial spheroid as estimated by BS, use the outer solution to
Poisson's equation for the bar, and assume that the non-axisymmetric
perturbation is dominated by its quadrupole term.  In order to compute
the velocity signature near both the ILR and OLR, we need $dU_2(r)/dr$
(cf. \ref{expcoef} and \ref{epiexp}).  One finds
\begin{equation} \label{Model}
	{dU_2(r)\over dr}=\cases{
		\displaystyle\epsilon {V_{rot}^2\over R_b} {2 r/R_b\over
		1 + (r/R_b)^5}&		bar;\cr\noalign{\vskip14pt}
		\displaystyle\epsilon {3\over4}{V_{rot}^2\over R_b}\left(r\over
		R_b\right)^\gamma&	spheroid,\cr
		}
\end{equation}
where $\epsilon$ is the strength of the bar relative to the
axisymmetric restoring force at the characteristic radius $R_b$.  BS
estimate $\gamma$ to be $2\le-\gamma\le2.5$.  The units and choice of
parameters parameters for these models are listed in Table 1.  A given
pattern speed $\Omega_b$ determines the radial location of the
resonance.  However, as the bar evolves, the pattern speed may
change and therefore the locations of the resonances may change.  We
describe the initial and final locations as $R_i$ and $R_f$.  The
parameters of the models we will discuss are presented in Table 2.


\def\sround#1{\hbox to2cm{\hfil#1\hfil}}

\begin{table}[t]
\caption{Model parameters}
\begin{tabular}{lccl}
Model	&	Parameter		& Value		& Comment \\
\tableline
Galaxy\tablenotemark{a} \\ \tableline
&	$R_{LSR}$			& 1.0		&	radius of LSR \\
&	$V_{rot}$		& 1.0		&	speed of LSR (flat rotation curve)\\
&	$\sigma_r$		& 0.1		&	radial
velocity dispersion in units of $V_{rot}$ \\
\tableline
Bar at OLR \\ \tableline
&	$R_b$			& 0.3		&	characteristic
radius of bar \\
&	$\epsilon$		& 0.2\tablenotemark{b}		&	ratio
non-axisymmtric to Galaxy force at bar end \\
&	$R_{OLR}$		& 0.64		&	location of resonance ($\approx10.4\kpc$)\\
\tableline
Spheroid at ILR \\ \tableline
&	$R_b$			& 1.0		&	spheroid scale factor \\
&	$\epsilon$		& 0.02		&	ratio of
non-axisymmetric to Galaxy force at LSR \\
&	$\gamma$		& $-2.0$		&	exponent of
quadrupole force powerlaw \\
&	$R_{ILR}$		& 1.3		&	location of resonance ($\approx5.1\kpc$)\\
\tablenotetext{a}{defines sysem of units}
\tablenotetext{b}{ratio of non-axisymmetric to
axisymmetric force at the solar circle is $1\%$}
\end{tabular}
\end{table}


\def\sround#1{\hbox to2cm{\hfil#1\hfil}}

\begin{table}[t]
\caption{Run parameters}
\begin{tabular}{cccc}
Run		& Resonance	&
\sround{$R_i$\tablenotemark{a}}	&
\sround{$R_f$\tablenotemark{b}} \\
\tableline
A&	ILR&	1.3&	1.3	\\
B&	   &	1.1&	1.3	\\
C&	   &	1.0&	1.3	\\
D&	   &	0.9&	1.3	\\
\tableline
I&	 OLR&	0.625&	0.625	\\
J&	    &	0.625&	0.75	\\
K&	    &	0.625&	0.875	\\
L&	    &	0.625&	1.0	\\
\end{tabular}
\tablenotetext{a}{initial resonance location}
\tablenotetext{b}{final resonance location}
\end{table}

\subsection{Evolution of an ensemble}

The goal is to compute the velocity along any given line of sight
incorporating effects induced by the bar.  The previous section (\S2)
outlines the evolution of a particular orbit.  In order to compute the
line-of-sight velocity at a particular point we need to average over
the entire evolved ensemble given by the initial distribution
function.  Here, we assume a Schwarzschild distribution with a
velocity dispersion $\sigma_r$.  Unfortunately, since the behavior
near resonance is non-linear, a closed form solution is not possible.
However as described in Appendix A, the post-evolution actions may be
determined as a function of the original actions using a simple
look-up table.

For discussion, we will consider initial ensembles described by a
constant value of $I_2$ and a distribution of $I_1$ consistent with
the Schwarzschild distribution.  For each $I_1$, all phases $w_1$ and
$w_2$ are equally represented. In the epicyclic limit, these ensembles
may be described by the initial value of the guiding center.  Note
that individual members of the post-evolution ensemble may be
distributed in guiding center radii and therefore ensemble averages do
not strictly represent the value at a point in space.  Also, a local
patch of the disk contains orbits from a distribution of guiding
center radii.  Nonetheless, the ensemble evolution gives some
indication of the expected kinematic signatures since the epicyclic
amplitudes are relatively small.  In addition, this definition of an
ensemble does greatly simplify the calculation and provides insight.
Although a tractable extension of the results presented here, the
derivation of the space-localized evolved distribution requires models
for a large fraction of the entire disk and gives less insight into
the evolutionary mechanism.  Such large-scale models will be useful
for comparing with a large spatially distributed kinematic data set
(see
\S4.3).

The radial and tangential velocities of the ensemble observed
along a line of sight at angle $\phi$ to the bar may then be written:
\begin{eqnarray}
	V_r & = & \Biggl\langle
		\int d J_f \int {d w_f\over2\pi} f(I_1, I_2)
			\left(\pm\Omega_1(R)a\cos w_1\right)
			\delta\left(w_s - w_s(w_1,w_2,\phi)\right)
		\Biggl\rangle,	\label{velrad}		\\
	V_t & = & \Biggl\langle
		\int d J_f \int {d w_f\over2\pi} f(I_1, I_2)
			\left(V_{rot}\pm\Omega_2(R)a\sin w_1\right)
			\delta\left(w_s - w_s(w_1,w_2,\phi)\right)
		\Biggl\rangle,  \label{veltan}
\end{eqnarray}
where $\pm$ is for ILR and OLR respectively, $\langle\,\rangle$
indicates the ensemble average at fixed $I_2$, and $w_s(w_1, w_2,
\phi)$ is the value of $w_s$ for a given $w_1$ and $w_2$ from the
look-up table.  Similar expressions may be derived for the velocity
dispersions, $\sigma_t = \sqrt{V_r^2  - (V_r)^2}$ and $\sigma_t =
\sqrt{V_t^2  - (V_t)^2}$, where
\begin{eqnarray}
	V_r^2 & = & \Biggl\langle
		\int d J_f \int {d w_f\over2\pi} f(I_1, I_2)
			\left(\Omega_1(R)a\cos w_1\right)^2
			\delta\left(w_s - w_s(w_1,w_2,\phi)\right)
		\Biggl\rangle,	\label{velrad2}		\\
	V_t^2 & = & \Biggl\langle
		\int d J_f \int {d w_f\over2\pi} f(I_1, I_2)
			\left(V_{rot}\pm\Omega_1(R)a\sin w_1\right)^2
			\delta\left(w_s - w_s(w_1,w_2,\phi)\right)
		\Biggl\rangle.	\label{veltan2}
\end{eqnarray}
Initially with no perturbation, the distribution is independent of
$w_s$ and therefore $V_r=0, V_t=V_{rot}$.  The post-evolution
trajectories depend on the initial phase but if the original
distribution is phase mixed than the new distribution should be phase
mixed for sufficiently slow evolution,
$\left|{{\dot\omega}_b\over\Omega_b^2}\right|$.  Therefore, the
distribution of $w_1$ and $w_2$ for the final orbit may then be
derived by sampling the final orbit at equal time intervals or,
equivalently, by assuming an flat distribution in the angle conjugate
to the post-evolution action:
\begin{equation}
	J_{final} = {1\over2\pi}\oint d w_s J_s \nonumber
\end{equation}

In order to determine the observable quantities (eqns.
\ref{velrad}--\ref{veltan2}), we must convert back to local variables.
Near the ILR or OLR, a librating orbit will still have a well-defined
guiding center trajectory instantaneously, but the apses will be
confined to a small range of position angles.  If we observe in a
frame rotating at the pattern speed, the trajectories will be
streaming forward (backward) near the ILR (OLR) on almost closed
orbits.  However, as long as the post-resonant epicyclic amplitude
remains relatively small, the trajectory is instantaneously close to a
valid epicycle.

\subsection{Velocity signature near resonance}

Using equations (\ref{velrad}) and (\ref{veltan}), we may predict the
line-of-sight velocity near resonance for any given scenario.  We will
be begin with a detailed investigation of the signature near ILR and
discuss specific scenarios below.

\subsubsection{Results at particular radii for a bar with constant
pattern speed}

The line-of-sight velocity and dispersion in an inertial frame fixed
at the Galactic center for Model A (see Table 1) are summarized in
Figures
\ref{fig:3n}--\ref{fig:5n}.  The angle $\phi$ describes
difference between the bar's position angle and the viewing
angle.\footnote{Angles and rotations are defined in the mathematical
sense, as if viewing from the SGP.}\ The guiding centers for the
ensembles are chosen for a range of discrete values of $R$.  Figure
\ref{fig:3n} (\ref{fig:4n}) describes the velocity signature for
ensembles with guiding centers inside (outside) $R_{res}$.  The sign
of the changes are as expected:

\begin{enumerate}
\item $R < R_{res}$. \quad Inside of ILR, the apses will tend
to be antialigned with the bar and orbits drift forward in the bar's
frame; therefore, in the first quadrant relative to the bar the mean
outward velocity will be positive.

\item $R > R_{res}$. \quad Outside of ILR
the apses will tend to be aligned with the bar and therefore, in the
first quadrant relative to the bar the mean outward velocity will be
negative.
\end{enumerate}

\noindent The large values of $V_r$ at small $R$ ($R\lesssim0.6$) are
not a significant feature of the resonance but due to the increasing
amplitude of the spheroid quadrupole (cf. eq. \ref{Model}).
Far outside the $R_{res}$, $V_r$ drops quickly (cf. Fig.
\ref{fig:4n}) due to the decreasing spheroid strength.

\subsubsection{Capture into libration and critical-crossing trajectories}

The picture is more complex near resonance as seen in Figure
\ref{fig:5n}.  The ``kick'' received from the evolving bar as the
orbit evolves through the critical trajectory accounts for the large
amplitude of the curves near $R\approx1.4$ even though these
trajectories are initially outside the resonance.  A significant
population of librating orbits causes the oscillation in $V_r$ with
$\phi$.

Figures \ref{fig:6a} and \ref{fig:6b} show the fractional contribution
to the line-of-sight velocities from non-critical, critical and
librating trajectories for $R=0.8$ and 1.4 (cf.  Figs. \ref{fig:3n}
and \ref{fig:5n}).  For $R=0.8$ few orbits have passed through the
critical trajectory and $15\%$ are in libration.  For $R=1.4$ nearly
all trajectories have both passed through the critical trajectory as
the potential evolved and are in libration.  For $R\lesssim0.9$ or
$R\gtrsim1.6$, no trajectories will have passed through the critical one
and received a jump in action; those at small $R$ will have a large
fraction of trajectories in libration (cf. Fig. 2) and those at large
$R$ will have none.\footnote{For an example of the number fraction of
orbits for each ensemble at $R$ that have crossed the critical
trajectory and/or been trapped into libration, see Fig.
\ref{fig:frac}.}

The initial ensemble has orbits evenly distributed in radial and
azimuthal phases and velocity dispersion or epicyclic energy chosen
according to the Schwarzschild distribution.  However, the response
for critical-crossing trajectories depends on the slow phase, $w_s$,
at the crossing.  To better understand the spatial morphology of these
orbits after the bar has evolved, we may select an ensemble of orbits
all with the same epicyclic energy $\sigma_r^2$ and average over all
phase but at fixed $w_s$ to get a ``mean'' orbit.  This may be thought
of as a time-averaged trace of the particular orbit at the final
(fixed) bar strength and frequency.  Figure \ref{fig:sorbAAA} shows
the mean orbits computed from 10 initially equally spaced $w_s$ for
$R=0.6, 0.9, 1.1, 1.3, 1.5$.  The bar major axis is along $\phi=0$.
For all but $R=1.3$, the mean orbits are nearly phase independent and
the 10 trajectories are simultaneous.  For $R\approx1.3$, capture into
libration depends on initial phase (cf.  Fig. \ref{fig:6b}) resulting
in different guiding center radii.  The librating orbits are captured
into the antialigned orientation and have the largest epicyclic
amplitudes in Figure \ref{fig:sorbAAA}.  The $R=0.6$ orbits are also
all in libration which gives them more elongation than, say, the case
with $R=0.9$ which has no librating orbits.

\subsubsection{Results for a bar with decreasing pattern speed}

Figure \ref{fig:sorbBBB} shows the mean orbits for Model B whose
pattern speed decreases by 17\% (see Table 1).  Since the fraction of
orbits that are captured into libration and cross the critical
trajectory depends on the evolutionary history and not simply the
final state, we find quantitatively different results.  The guiding
center $R=0.9$ is broken up into two distinct groups. The inner
``boxy'' mean trajectories have been captured into libration and the
outer trajectories have not. All have crossed the critical trajectory.
The librating trajectories are antialigned with the bar and the
rotating trajectories are aligned with the bar.  Guiding centers with
$R\gtrsim1.7$, which are non-critical and in rotation, only depend on the
final value of $\Omega_b$ and not on its history.  Overall, the
changing pattern speed in Model B has increased the fraction of
libration and critical trajectory-crossing orbits over Model A.  As we
will see below, these differences are reflected in observable
kinematic signatures and may be used to limit evolutionary hypotheses.

\subsubsection{Expectations at a specified position angle}

The dispersion and velocity profiles along particular lines of sight
are shown in Figures \ref{fig:vrsAAA} and \ref{fig:vrsBBB} for Models
A and B.  These plots represent cuts at constant $\phi$ through (e.g.)
Figures \ref{fig:3n}--\ref{fig:5n}.  We select $\phi=-45^\circ$,
anticipating comparison with a rotating spheroid whose position angle
is in the 4th quadrant in Galactic longitude.  Note that the overall
amplitude and radial breadth of the feature is larger in the case of
Model B than A.  This trend is found for most models; the larger the
variation of the model parameters with time, the more change in the
trajectories.  Therefore a model which both grows in strength and
evolves in pattern speed shows a stronger response (Model B) than one
which only grows in strength (Model A).  Although the graph of $V_r$
vs.  $\phi$ is smooth for individual trajectories, after evolution the
curves may overlap, especially near resonance, and produce a distinct
signature along particular lines of sight.  For example the dip in
$V_r$ at $R=1.1$ in Figure
\ref{fig:vrsBBB} is caused by the superposition of librating and
non-librating orbits (cf. Fig. \ref{fig:sorbBBB}).  The response at
constant perturbation strength increases dramatically with
$\Delta\Omega_b$; some of the orbits near ILR in Models C and D are
perturbed so strongly after evolution the epicyclic approximation is
invalid.  The observed lack of anomalously large velocities and
dispersions expected from these distorted orbits near the solar
position would suggest that a rotating spheroid must either have an
extremely stable pattern speed or small amplitude.

Figure \ref{fig:sepAAA} shows the relative contribution to the
line-of-sight velocity orbits which have become critical or captured
into libration in Model A.  A large fraction of all the orbits between
1 and 1.5 passed through the critical trajectory as the rotating
perturbation grows in strength.  We see that the fraction of librating
orbits follows the run of velocity dispersion in Figure
\ref{fig:vrsAAA}.  At small $R$, the relative strength of the
perturbation increases and the orbits find themselves in the libration
zone without having been critical.

To summarize, the strongest perturbations are caused by the non-linear
response and are in a band about the formal location of the resonance.

\newpage

\section{Application to the Galaxy} \label{sec:galaxy}
Resonances (ILR or OLR) may lead to pronounced kinematic signatures in
the line-of-sight velocity and velocity dispersion (\S3).  In this
section, we discuss two cases in detail: a triaxial spheroid and
standard stellar bar.

\subsection{ILR models and implications for a rotating triaxial spheroid}

In order to explain the motion of the LSR inferred from the asymmetry
in the HI $l$-$v$ diagram, Blitz and Spergel (1991) postulate a
rotating triaxial spheroid with $\Omega_b=6\kms/\kpc$.  The parameters
in Models A---D are chosen in accord with these parameters, although
there was no attempt to tune the models to reproduce a particular LSR
velocity.

Each of the models describes a possible evolutionary trend in the
rotating spheroid.  Model A with constant pattern speed shows a
discontinuity in $V_r$ around the resonant radius at $R=1.3$
corresponding to $10.4\kpc$ scaled to Galactic units (Fig.
\ref{fig:data}).  This model would imply an outward motion of the LSR
of about $20\kms$ (similar to BS's estimate of $14\kms$ and differing
because of the nonlinear response).  Outside of $R\approx14\kpc$,
$V_r$ would be unchanged and therefore the outer Galaxy would appear
to be approaching the Sun.  Note that there is a strong peak predicted
in the line-of-sight velocity dispersion about the resonance location.

Model B--D have decreasing pattern speeds.  In the case of (e.g.)
Model B, the resonance moves from $R=8.8\kpc$ to its final position at
$R=10.4\kpc$.  A larger measure of orbits have been perturbed by the
resonance giving a broad increase in velocity dispersion and smearing
and shifting the line-of-sight velocity profile as seen in Figure
\ref{fig:vrsBBB}.  The amplitude of $V_r$ is a factor of 2 larger in
this case.  An evolving pattern speed produces an observable signature
with smaller spheroid amplitudes than assumed by BS, suggesting that
velocity kinematics may be an even more sensitive probe of asymmetry
than previously assumed.

Metzger and Schechter (1992) found that carbon stars in the direction
of the Galactic anticenter appear to be systematically receding from
the LSR (cf. Figure \ref{fig:data}).  This is also consistent with
the K-giant data from Lewis \& Freeman (1989).  A naive interpretation
suggests that the net stellar motion opposes the net gas motion.  On
the other hand, the motions might be better explained by a spheroid
with $\phi\approx+45^\circ$ giving rise to an inward LSR motion.
However, this counters the original motivation by Blitz and Spergel
(1992) to explain the asymmetry in the HI $l$--$v$ diagrams.
Regardless, rotating spheroid models predict a strong jump in velocity
dispersion near the OLR and provides a method to predict (or limit)
their amplitude.

\subsection{Implication for the Inner Bar}

As emphasized in previous sections, ILR ($l_1=-1, l_2=2$) is closely
related to OLR ($l_1=1, l_2=2$).  The governing equations
(\ref{hamres}, \ref{hamlr}) are identical and the physical
explanations above apply, with appropriate changes in sign.

Consistent with Weinberg (1992), we assume a bar ending at
corotation at $3\kpc$, major-axis position angle of $45^\circ$, and
quadrupole strength of 20\% of the axisymmetric background force at
the end of the bar, corresponding to a strong stellar bar.  This gives
an OLR at $\approx5\kpc$ or $R=0.625$ in model units.  Four models, I--L,
consistent with these parameters are described in Table 1.  Model I
has a static pattern speed and all others have a decreasing pattern
speed.  Figure \ref{fig:vrsII} shows Model I.  The LSR motion is
unchanged by the perturbation but the line-of-sight velocity increases
toward the OLR with a factor of $\sim2.5$ jump in line-of-sight
velocity dispersion about the OLR.

We expect a moderate bar to lose angular momentum as it evolves
(\cite{wei85,LiCa:91,her92}).  If this torque causes the pattern speed
to decrease, the position of the OLR will increase, causing a large
increase in velocity dispersion in the vicinity of OLR ($\sim5\kpc$)
with little change in the net radial motion.  Figure \ref{fig:vrsJJ}
shows the results of Model J whose OLR moves from $5$ to $6\kpc$.  The
radial velocity peak broadens and decreases in amplitude by a factor
of 2 while the dispersion broadens and increases by a factor of two.
Figure \ref{fig:lf} shows Model J scaled to $R_{LSR}=8\kpc$
$V_{rot}=220\kms$ (solid line and open circles) together with the
Lewis \& Freeman velocities (open squares and error bars); the
observations suggest the predicted signature for the stellar bar.
There has been no attempt to fit the model to the data other than
selecting Model J from I--L.

Radakrishnan \& Sarma (1980) estimate a radial dispersion of gas
clumps in the direction of the center of $5\kms$ and a systemic
velocity of $<1\kms$ based on the HI absorption spectrum of Sgr A.
Although one expects the gas dispersion to be lower, in general, than
the stellar velocity dispersion, the low systemic velocity is
consistent with the predicted and observed trends in stellar
kinematics: the perturbation to the line-of-sight velocity is small
while the large dispersion is due to intersecting librating orbits.

\subsection{Future Observations and Analyses}

For an axisymmetric Galaxy, we expect $V_r=0$ for all $R$ in the
direction of the center or anticenter, and since these measurements
are natural diagnostics for asymmetry, they have received the most
attention.  However, the models described in previous sections predict
distinct features in the velocity dispersion as well as the systemic
velocity because of jumps in action and capture into libration near
the resonances (cf.  Figs. \ref{fig:3n}--\ref{fig:5n} and
\ref{fig:sorbAAA}--\ref{fig:sorbBBB}).  These features allow for
explicit testing of the various rotating spheroid and bar models using
a spatially distributed sample of kinematic tracers.  The theoretical
models described above (Table 1) are easy to compute, allowing a wide
variety of cases to be tested.  Existing K-giant, carbon star, Mira
and Cepheid variable data may permit precise testing of the various
bar hypotheses using the standard statistical estimators (e.g.
likelihood) and models constructed for full annuli of the Galactic
disk rather than individual guiding center radii.  This work is in
progress.

\section{Summary}

The major conclusions are as follows:
\begin{enumerate}
\item A rotating pattern, such as a bar or spheroid, causes a
distinctive stellar kinematic signature near primary resonances (OLR
and ILR).  The amplitude is larger than one would predict from the
mean field of the static potential perturbation alone because of
nonlinearity of the resonant response.  These resonances may be far
from the peak density of the pattern and relatively nearby the solar
position for both a triaxial spheroid and Galactic bar, raising the
possibility for direct observation.
\item Near OLR or ILR, a fraction of the trajectories for a given
guiding center will be trapped into libration; the probability of
trapping depends on phase.  Overlapping librating and non-librating
trajectories increase the velocity dispersion.  Velocity dispersion
measurements may be as useful as the systemic velocities in testing
for the existence of a rotating disturbance.  For example, a moderate
strength Galactic bar ending at $3\kpc$ may easily produce an
increased velocity dispersion at OLR ($\sim5\kpc$) of $50\kms$ or
larger.
\item The details of the signature depend on the evolutionary history
of the bar; a changing pattern speed or perturbation strength change
the details of the response.  The fraction of orbits both trapped into
libration and strongly perturbed by passing through resonance depends
sensitively on the change in pattern speed; a 20\% change in pattern
speed may increase the velocity dispersion by a factor of 2--3.
Therefore, kinematic data may be used to probe the evolutionary
history as well as the present state of the Galaxy.
\item Blitz \& Spergel (1991) suggest that the LSR motion may be explained by a
large-scale rotating spheroid.  It has been recently pointed out
(\cite{met92}) that the stellar kinematics are inconsistent with this
simple picture.  Various possible evolutionary scenarios are explored
but none allow the gas and stellar kinematics to be simply understood
with the proposed rotating spheroid model.  In addition to LSR motion,
a rotating non-axisymmetric spheroid will produce a velocity
dispersion increase near ILR which should be observable.
\item I have predicted the kinematic signature that might be found for
a Galactic bar with semimajor axis of $3\kpc$ such as inferred by
Weinberg (1992).  We expect a moderate bar to lose angular momentum as
it evolves (\cite{wei85,LiCa:91,her92}).  If this torque causes the pattern
speed to decrease, the position of the OLR will increase.  This will
cause a large increase in velocity dispersion in the vicinity of OLR
($\sim5\kpc$) with little change in the net radial motion.  Such a
signature is suggested by K-giant velocity data (\cite{lew89}) and
consistent with HI gas data (\cite{RaSa:80}).
\end{enumerate}

\acknowledgments

I thank Leo Blitz and Dave Spergel for discussions and Susan Kleinmann
for a critical reading.  This work was supported in part by NASA grant
NAG 5-1999.

\newpage
\appendix
\section{Computational Method} \label{app:comp}

Here, we develop and efficient computational form for equation
(\ref{hamres}).  In the cases of interest, $l_1=\pm1$ and the
potential coefficient $W^2_{l_1\,l_2}\propto a\propto\sqrt{I_1}$.  In
terms of actions, $a\propto\sqrt{J_s + J_f}$ which makes equation
(\ref{hamres}) numerically inconvenient.  However, it is
straightforward to make a canonical transformation to a new set of
variables where $\cJ\equiv J_s + J_f$ keeping $J_f$ constant.  The
transformed Hamiltonian becomes:
\begin{equation} \label{hamlr}
	{\bar H}^\prime = {1\over2}G\cJ^2 - G\cJ_o\cJ
	-{\dot\Omega}_b(t-t_{res})\cJ - F\sqrt{\cJ}\cos(\theta),
\end{equation}
where $G\equiv\left.{\partial^2H_o/\partial J^2_s}\right|_{res}$,
$F\equiv{dU_2/ dr}|_{res}$, $\cJ_o\equiv J_{s,res}+J_f$, $\theta=w_s+\pi/2$,
and constant terms have been dropped.  Both the coefficients of the
$\cJ$ and $\sqrt{\cJ}$ terms are in general time-dependent in our
model.  However, as long as $\dot\Omega$ and $\dot F$ are very small,
the trajectory remains near a solution with fixed coefficients for
many dynamical times.

Numerical solution to the equations of motion generated by equation
(\ref{hamlr}) are further complicated by the $1/\sqrt{\cJ}$ term in
the equations of motion.  This singularity in the coordinate system
$(I,\theta)$ may be removed by a transformation to a rectangular
coordinate system, $(x, y)$.  This may be affected with the generating
function $S_1(y,\theta)=y^2\cot\theta/2$ (1st kind, \cite{gol50})
which gives the transformation $x=\sqrt{2\cJ}\cos\theta,
y=\sqrt{2\cJ}\sin\theta$.  The new Hamiltonian is then:
\begin{equation} \label{hamlrc}
 	{\bar H}^{\prime\prime} = {1\over8}G(x^2+y^2)^2
-{1\over2}\left[G\cJ_o-{\dot\Omega}_b(t-t_{res})\right](x^2+y^2) -
{F\over\sqrt{2}}f(t) x.
\end{equation}
The function $f(t)$ is chosen to slowly and smoothly vary from $0$ at
$t=T_{min}$ to $1$ for $t>T_{max}$.  To compute the response of an
orbit to the adiabatically growing potential, $\cJ_o$ is first
computed from the given initial values of $I_1$ and $I_2$.  The
equations of motion are then integrated until $t>T_{max}$ and the new
value $J_s$ is computed from $\cJ$.  Since we are only interested in
the ensemble average, it is sufficient to tabulate $\cJ(\cJ_o,\theta)$
for given values of $G$ and $F$.  For a logarithmic background
potential, the quantity $G=-2(2\pm\sqrt{2})/R_{res}^2$ for OLR and ILR
at $R_{res}$ respectively.  In practice, I choose
\begin{equation}
	f(t) = {1\over2}\left[ 1 + \hbox{\rm
erf}\left(t-T_{max}\over\tau\right)\right]
\end{equation}
and increase $\tau$ with appropriate choice of $T_{min}$ and $T_{max}$
to verify that the evolution is adiabatic.

Away from the resonance the solutions will scale with $F$.  However,
near the resonance, measure of orbits that cross the critical
trajectory will depend on $F$.  In fact, as long as the orbit has not
become critical or captured into libration then the response should
scale linearly and this has been verified numerically.  Here, we are
explicitly interested in the nonlinear effects of the resonances which
requires us to solve the model for each value of $F$ and $G$ of
interest.

\section{Evolution Through Resonance: technical discussion} \label{app:tech}

Since the orbit morphology near resonance is governed by equation
(\ref{hamlr}) or (\ref{hamlrc}), the dynamical consequences by be
inferred from one-dimensional solutions directly.  To reiterate, the
appearance of the orbit changes with the evolving bar but the actions
are adiabatically invariant except near critical trajectories.  As the
trajectory passes through the critical trajectory, the action may
change discontinuously by a finite amount but then remain invariant
thereafter as in the case of the pendulum.  The critical trajectory,
divides the phase plane into distinct regions.  The probability of a
particular trajectory entering one or the other depends on the phase
at the critical trajectory.  Hamiltonian functions of the form
equation (\ref{hamlr}) have been studied in some detail (e.g.
\cite{hen83},
\cite{lic83}), and here we give an example of these graphical methods
for Model I (cf. Tables 1 and 2).

Figure~\ref{fig:1} shows some possible trajectories in this
Hamiltonian where the $x$ and $y$ axes are $\sqrt{2\cJ}\cos\theta$ and
$\sqrt{2\cJ}\sin\theta$ respectively.  Physically, the radius
$\sqrt{2\cJ}$ is proportional to the epicyclic radius and the phase
angle $\theta$ describes the rotation or libration about the
resonance.  The critical points (equilibria) have $\theta=0,\pi$.  In
the case of Figure~\ref{fig:1}, the only (stable) critical point has
$\sqrt{2\cJ}=-0.156$ and no other critical trajectory.  The
corresponding trajectory is a closed oval in the frame rotating with
the pattern and often is called a {\it periodic orbit.}\/ Around this
critical point the orbits are in libration; the angle of the apocenter
is confined to a range smaller than $[0,2\pi]$.  For sufficiently
large $\sqrt{2\cJ}$, the trajectory includes the origin and the
trajectory circulates: the angle of the apocenter precesses through
$2\pi$ radians.  As the strength of the bar grows the critical point
shifts further to the left and the measure of librating trajectories
grow.  A librating trajectory has its phase at apocenter ``trapped''
opposite to the position angle of the bar in this case.

In the case of Figure~\ref{fig:2}, the guiding center is inside the
resonance initially and there are three critical points, one at the
center of the librating orbit (stable), one at the center of the tiny
loop (stable), and one at the $\times$ (unstable).  The critical
trajectory, which terminates at the unstable point, divides the phase
space into three regions.  As the potential grows, a trajectory may
switch region, ending up either in rotation or trapped into libration.
Here, the trapping causes a jump in epicyclic amplitude.  and changes
the apparent line-of-sight velocity signature as a function of
viewpoint.  Note that the fraction of librating orbits is not
symmetric on either side of the resonance due to the topological
differences.  In addition, trajectories in the small loop will librate
with small amplitude along the position angle of the bar while
trajectories like the kidney-shaped one shown exhibit large amplitude
librations orthogonal to the position angle of the bar.  In the case
of models with changing pattern speeds, the phase space topologies are
similar to Figures \ref{fig:1}--\ref{fig:2} although the size of the
librating regions, for example, may change.  Diagrams of this sort
allow quantitative comparison of evolutionary scenarios without
constructing observable quantities.

\clearpage
\newpage

\begin{figure}
\caption{\label{fig:trap} Librating orbit near ILR after bar evolution;
the strength and pattern speed are constant.  The quadrupole force is
2\% of the axisymmetric force at $R=1$. The resonance is at $R=1.3$
and the initial guiding center is at $R=1.35$.}
\end{figure}

\begin{figure}
\caption{\label{fig:traporb} Graph of the orbit described in Fig.
\protect{\ref{fig:trap}} as seen from the bar frame.  The first (last) 30
time units are shown as a solid (dashed) line.  The bar is oriented
along the $x$-axis.}
\caption{\label{fig:intlorb} Graph of the orbit as seen from the
inertial frame.  The bar rotates in this frame at frequency
$\Omega_b$.}
\end{figure}

\begin{figure}
\caption{\label{fig:untrap} As in Fig. \protect{\ref{fig:trap}} but
for a rotating orbit with initial guiding center is at $R=1.0$.}
\caption{\label{fig:untrpf} As in Fig. \protect{\ref{fig:trap}} but
for a rotating orbit with initial guiding center is at $R=2.5$,
outside the resonance.  Note that the run of $\phi_{apo}$ is
retrograde with time in this case (outside resonance) and prograde in
Fig. \protect{\ref{fig:untrap}} (inside resonance).}
\end{figure}

\begin{figure}
\caption{\label{fig:3n} The line-of-sight radial velocity for
different guiding center radii.  The quantity $V_r$ is relative to the
galaxy's inertial frame (not the LSR!) and the angle $\phi$ describes
the angle between the bar's position and the line of sight.  Each
curve represents the mean velocity for an ensemble of orbits with a
Schwarzschild velocity distribution with radial dispersion of
$0.1V_{rot}$ and initial guiding center trajectory $R$.  The resonant
radius is $R_{res}=1.3$ (coresponding to $10.4\kpc$ if $R_{LSR}=8\kpc$).}
\caption{\label{fig:4n} As in Fig. \protect{\ref{fig:3n}} but shows
large guiding center radii ($R>R_{res}$).}
\end{figure}

\begin{figure}
\caption{\label{fig:5n} As in Fig. \protect{\ref{fig:3n}} but shows
guiding center radii near the resonance ($R=1.3$).}
\end{figure}

\begin{figure}
\caption{\label{fig:6a} Shows the relative contribution to $V_r$ for
$R=0.8$ from orbits which have received a kick in action
(critical-crossing) and from those in libration.  The total is shown
as a heavy solid line.}
\end{figure}

\begin{figure}
\caption{\label{fig:6b} As in Fig. \protect{\ref{fig:6a}} but for $R=1.4$.
Note that
near the resonance (at $R\approx1.3$), nearly all trajectories are
homoclinic crossing.}
\end{figure}

\begin{figure}
\caption{\label{fig:sorbAAA} Mean trajectories for Model A and initial guiding
center radii as labeled.  Ten different initial phases are shown for
each guiding center.  For all but $R=1.3$, the curves are independent
of initial phase and are coincident.  The bar position angle is
$0^\circ$.}
\end{figure}

\begin{figure}
\caption{\label{fig:sorbBBB} As in Fig. \protect{\ref{fig:sorbBBB}} but
for Model B.  For all cases but $R=1.7$, the curves are no longer phase
independent.  In the case of initial guiding center $R=1.1$, the
trapping depends on initial phase (the orbits elongated along
$\phi=90^\circ$ and $270^\circ$ are trapped.)}
\end{figure}

\begin{figure}
\caption{\label{fig:vrsAAA} Line-of-sight radial velocity $V_r$, and dispersion
$\sigma_r$ for Model A at $\phi=-45^\circ$.  The symbols show
evaluations of $V_r$ and may be thought of as a cut at $\phi$ in
Figures
\protect{\ref{fig:3n}}--\protect{\ref{fig:5n}}; the connecting lines
are provided to guide the eye.}
\caption{\label{fig:vrsBBB} As in Fig. \protect{\ref{fig:vrsAAA}} but
for Model B.  The dip at $R=1.1$ is caused by the superposition of
both trapped and untrapped orbits.}
\end{figure}

\begin{figure}
\caption{\label{fig:sepAAA} As in the lower panel of Fig.
\protect{\ref{fig:vrsAAA}} but
showing the individual contributions to $V_r$ from post-critical
orbits (solid) and librating orbits (dashed).  The total (dotted) is
shown for comparison.  The resonance is at $R=1.3$.}
\end{figure}

\begin{figure}
\caption{\label{fig:frac} Shows the fraction of orbits that have
crossed the critical trajectory (solid) or have been trapped into
libration (dashed) for the ensemble shown in Fig.
\protect{\ref{fig:sepAAA}}.}
\end{figure}

\begin{figure}
\caption{\label{fig:vrsII} As in Fig. \protect{\ref{fig:vrsAAA}} but for
Model I.}
\caption{\label{fig:vrsJJ} As in Fig. \protect{\ref{fig:vrsAAA}} but for
Model J.}
\end{figure}

\begin{figure}
\caption{\label{fig:data} Comparison of Model A scaled to
$V_{rot}=220\kms$ and $R_{LSR}=8\kpc$ compared with Lewis and Freeman's
K-giant and Metzger and Schechter's carbon star velocity data.}
\end{figure}

\begin{figure}
\caption{\label{fig:lf} Lewis and Freeman's K-giant velocity
data compared with Model J.}
\end{figure}

\begin{figure}
\caption{\label{fig:1} Shows trajectories for a family of orbits with
guiding center radius corresponding to $0.7$ if the OLR radius is
$0.64$.  The resonance location is fixed and the bar strength
$|\epsilon|$ slowly grows to its maximum of 0.2. The quantity
$\protect\sqrt{2\cJ}$ is proportional to the epicylic radius $a$ so
small radii in the figure correspond to nearly circular orbits.  There
is only one critical point and no critical trajectory.  If the bar
were very weak, the trajectories would have constant $\cJ$ and be
circles in this plot.}
\end{figure}

\begin{figure}
\caption{\label{fig:2} As in Fig. \protect{\ref{fig:1}} but with
guiding center radius inside the OLR (corresponding to $0.475$).  In
this case there are three critical points (one at the center of each
small loop and one at the $\times$) and a critical trajectory.}
\end{figure}

\begin{figure}
\caption{\label{fig:3} Location of critical points for the models described in
Figs.~\protect{\ref{fig:1}} and \protect{\ref{fig:2}} in units of the
solar radius, $R=1$.  The vertical dotted line
shows the point at which the bifurcation occurs.  The two new critical
points appear at the point marked by the open dot.  The upper locus is
(dashed) is unstable.}
\end{figure}

\end{document}